\documentclass[journal=jctcce,manuscript=letter]{achemso}
\setkeys{acs}{articletitle = true}

\usepackage[version=3]{mhchem} 
\usepackage{subfigure}

\usepackage[T1]{fontenc}
\usepackage{graphicx}
\usepackage{caption}
\usepackage{xcolor} 
\newcommand{\red}[1]{\textcolor{black}{#1}}
\makeatletter
\usepackage{bm}
\usepackage[all]{nowidow}
\makeatother

\title{Determination of Density Functional Tight Binding Models for Cerium-containing Materials}

\author{Nir Goldman}
\affiliation{Physical and Life Sciences Directorate,
             Lawrence Livermore National Laboratory, Livermore, CA 94550 USA}
\email{ngoldman@llnl.gov}
\alsoaffiliation{Department of Chemical Engineering, University of California, Davis, California 95616, United States}

\author{Artem Samtsevych}
\affiliation{Fritz Haber Institute of the Max Planck Society, 14195 Berlin, Germany}

\author{Chiara Panosetti}
\affiliation{Fritz Haber Institute of the Max Planck Society, 14195 Berlin, Germany}
\begin{document}

\newpage
\begin{abstract}

We have developed Density Functional Tight Binding models for cerium \red{and cerium oxide} with generalized gradient approximation and hybrid functionals that accurately predict the electronic band structure of different \red{cerium polymorphs as well as the insulating oxide}.  We show that determination of a many-body repulsion energy with the Chebyshev Interaction Model for Simulation correctly predicts cerium allotropic energetic ordering \red{with a minimal training created from small-scale molecular dynamics calculations of a single phase, only.} \red{Our approach determines DFTB models for both electronic and material property predictions that retain a high degree of transferability.}

\end{abstract}

Density Functional Theory (DFT) calculations of materials with partially filled f-electron shells have a high degree of complexity due to the interplay between electron localization and hybridization, where changes in these interactions can lead to an abrupt volume collapse and phase changes.\cite{americium_1992, Gd_Y_collapse_2013, cerium_xray_2019} Cerium metal is a noteworthy example, where choice of exchange-correlation functional and basis set can determine whether the fcc-phase volume collapse from the lower density, ferromagnetic $\gamma$ phase to the higher density paramagnetic $\alpha$ phase can be observed.\cite{Scheffler_cerium_2016, Soderlind2025, cerium_dft_2026} \red{Oxides such as CeO$_2$ pose a significant challenge as well, where standard approaches such as the Generalized Gradient Approximation (GGA) generally underestimate electron correlations which results in the erroneous prediction of a metallic (vs. insulating) ground state. This can make DFT calculations particularly cumbersome, where either semi-empirical GGA+U approaches\cite{Hernandez_PuO2_2016, Aradi17, Stimac2025_ceo2} or hybrid functionals\cite{Pegg_UO2_NpO2, Pegg_PuO2} are required to enforce a band gap. Hybrid functionals have strong  appeal in that they can more accurately account for electronic exchange with less empiricism, though these approaches are between 10--10$^3$ times more computationally intensive than standard GGA\cite{functionals_2024}.} In addition, larger supercells can be required for molecular dynamics (MD) simulations,\cite{ReaxFF_PdH_2014, Oppelstrup_dPu_2025} calculation of defect properties,\cite{Goldman_SEQM_2023} or determination of charge carrier localization,\cite{polarons_tb_2000} which compounds the issue. Hence, determination of f-electron material properties would greatly benefit from methods with enhanced computational efficiency that can accurately describe electronic states with the accuracy of higher-order methods.  

In this regard, Density Functional Tight Binding (DFTB) is a semi-empirical quantum method derived directly from Kohn-Sham DFT that has proven to be an effective alternative for f-electron systems.\cite{Hourahine10, Aradi17, Goldman_DFTB_H_Pu, Goldman_PuO2H_2022} The formalism for DFTB with self-consistent charges (SCC) is discussed in detail elsewhere\cite{DFTB+_scc,Koskinen09,DFTB3,DFTB+,DFTB+_current,  DFTB+_2025}. Briefly, DFTB assumes neutral, spherically symmetric, atom-centered charge densities and performs a second-order Taylor expansion of the Kohn-Sham total energy in charge fluctuation. This results in the following total energy expression:

\begin{equation}
E_{\rm{DFTB}} = E_{\rm{BS}} + E_{\rm{Coul}} + E_{\rm{Rep}}.
\end{equation}

\noindent Here, $E_{\rm{BS}}$ corresponds to the band structure energy (determined from the approximate Kohn-Sham eigenstates), $E_{\rm{Coul}}$ is the charge fluctuation term (computed self-consistently), and $E_{\rm{Rep}}$ is the repulsive energy (which corresponds to ion-ion repulsion as well as Hartree and exchange-correlation double counting terms). Charge transfer is controlled through a Hubbard $U$ hardness parameter, determined \emph{ab initio} from the difference between the DFT computed ionization energy and electron affinity for an isolated atom. This value can be resolved per orbital angular momentum channel, accounting for different dispersion/localization for each orbital type. $E_{\rm{Rep}}$ is commonly determined via empirical function optimization, where parameters are fit to reproduce high-level quantum or experimental reference data.

The DFTB Hamiltonian matrix elements are determined from pre-tabulated Slater-Koster tables derived from DFT atomic dimer calculations with a user-specified functional and a minimal basis set. The onsite matrix elements are the free atom orbital energies and the two-center off-site terms are computed with both wavefunctions and electron density subjected to confining potentials. In general, the confining potentials are the most commonly tuned part of the electronic interactions in DFTB. Other parameters that factor into the DFTB Hamiltonian, i.e., the onsite orbital energies or the Hubbard $U$ charge transfer parameters, are usually determined completely \emph{ab initio} from free atom calculations using the uncompressed basis set and electron density.   \red{DFTB calculations thus provide an advantage over machine-learned MD potentials in yielding generally improved transferability as well as by producing information about the system's electron states through solving an approximate Kohn-Sham equation.} 

In this work, we demonstrate the creation of DFTB models for \red{f-electron materials,} starting from optimizing the confining potentials for the wavefunctions and electron density to determination of the repulsive energy. We choose to focus on cerium and cerium oxide (ceria) as model systems, as \red{cerium metal} has a rich solid phase diagram with allotropes at varying density that yield a wide range of electronic interactions, \red{while ceria serves as an example oxide that can be more accurately modeled via hybrid functional.} We have created \red{cerium metal} DFTB parameterizations with the Perdew-Burke-Enzerhof (PBE) GGA functional\cite{Perdew:1996} as well \red{parameterizations for both the metal and ceria} using the PBE0 hybrid functional\cite{PBE0}, which includes 25\% exact exchange. \red{To the best of our knowledge, our results represent the first semi-empirical quantum mechanical models parameterized to f-electron hybrid functional band structure results, specifically.} Our results illustrate the potential accuracy and transferability of DFTB for f-electron systems that could yield additional materials parameterizations of this type in the future.

Here, we choose to focus on nonmagnetic DFT and DFTB calculations only for both materials in order to simplify the DFTB optimization process. DFT calculations were performed using the Vienna ab initio Simulation Package (VASP)\cite{vasp,vasp2,vasp3}, with projector-augmented wave function (PAW) pseudopotentials\cite{Bloechl94,Kresse99}. For all of our calculations, we have used a planewave cutoff of 500~eV, Gaussian electron smearing with a width of 0.05~eV, and electron density convergence of $10^{-6}$~eV. Initial electron density optimizations for PBE band structure calculations were performed with a $20 \times 20 \times 20$ Monkhorst-Pack k-point mesh.\cite{Monkhorst76} PBE0 band structure calculations were initialized with a $9 \times 9 \times 9$ k-point mesh only due to their computational expense. Primitive cells for band structure calculations were determined by performing zero pressure optimizations from structures taken from the Materials Project\cite{materials_project} for each phase.

\begin{table}[!htp]
\caption{Decoding mask used in cerium DSKO optimizations. All band positions are shown relative to the uppermost valence band, which is defined as `0'.}
\label{tab:decoding mask}
\begin{tabular}{cc}
\hline
Special k-point  & Bands  \\
\hline
\hline
$\Gamma$ & $\left[-1, 0, 1, 4\right]$ \\
$X$ & $\left[-1, 0, 1\right]$ \\
$W$ & $\left[-1, 0, 1\right]$ \\
$K$ & $\left[-1, 0, 1\right]$ \\
$L$ & $\left[-1, 0, 1\right]$ \\
$U$ & $\left[-1, 0, 1\right]$ \\
\hline
\end{tabular}
\end{table}

All DFTB calculations were performed using the DFTB$+$ code\cite{DFTB+_2025} with SCC and an identical k-point mesh or set of lines as our DFT calculations. Band structure calculations were performed on primitive cells taken from our DFT optimizations. Electron populations were thermally smeared using the Mermin functional\cite{Mermin65} and a temperature of 600~K. The electron density was converged to $2.72 \times 10^{-5}$~eV ($10^{-6}$~au) for the results presented here. 

We have used the DFTB Slater-Koster Optimization (DSKO) code to perform global optimization searches for confining potential parameters,\cite{DSKO} where we have modified the code to allow for hybrid functional optimizations \red{(see the Supplementary Information for detailed code modifications).} DSKO computes an objective function for the band structure through a decoding mask, which is defined by energy differences between specified bands at special k-points. \red{We first focus on our cerium metal band structure optimizations.} Optimizations for PBE0 and PBE were performed using an identical decoding mask for both functionals (Table~\ref{tab:decoding mask}) on an $\alpha$-Ce (face centered cubic; fcc) primitive cell at both the DFT optimized geometry as well as an equally weighted primitive cell with lattice constants compressed by a factor of 0.98 in order to create a slightly diverse training set. Energies at selected k-points were chosen to include those from the valence band plus one band below and one band above, excluding the $\Gamma$-point where we also included the fourth energy above to help with predictions of high energy states. 

In all cases, we have optimized a Woods-Saxon confining potential\cite{WoodsSaxon} for the density as well as each of the s, p, d,  and f-orbitals separately. The p-orbitals in $\alpha$-Ce are unoccupied, which permits optimization of its onsite energy\cite{TDDFTB-ChIMES}. The DFTB$+$ code is currently unable to perform orbital resolved SCC calculations with hybrid functionals, though DSKO can still provide these values regardless of functional (through the skprogs code\cite{skprogs} for this work). Hence, PBE0 optimizations included a single Hubbard $U$ value, which essentially optimized to a mean-field value for all angular momentum channels in the system. In contrast, GGA functionals in DFTB$+$ allow for orbital resolved SCC models where each $U$ value was determined \emph{ab initio}. DSKO optimizations for both functionals were first performed with the particle swarm method\cite{PSO, DFTB_transferability_1} with 46 particles for 50 iterations, followed by fine tuning with Bayesian optimization\cite{BayesianOpt1, BayesianOpt2} for 80 iterations. Optimized parameters for both DFTB models are listed in the Supporting Information.

\begin{figure}[!htp]
\centering
\includegraphics[scale=0.25]{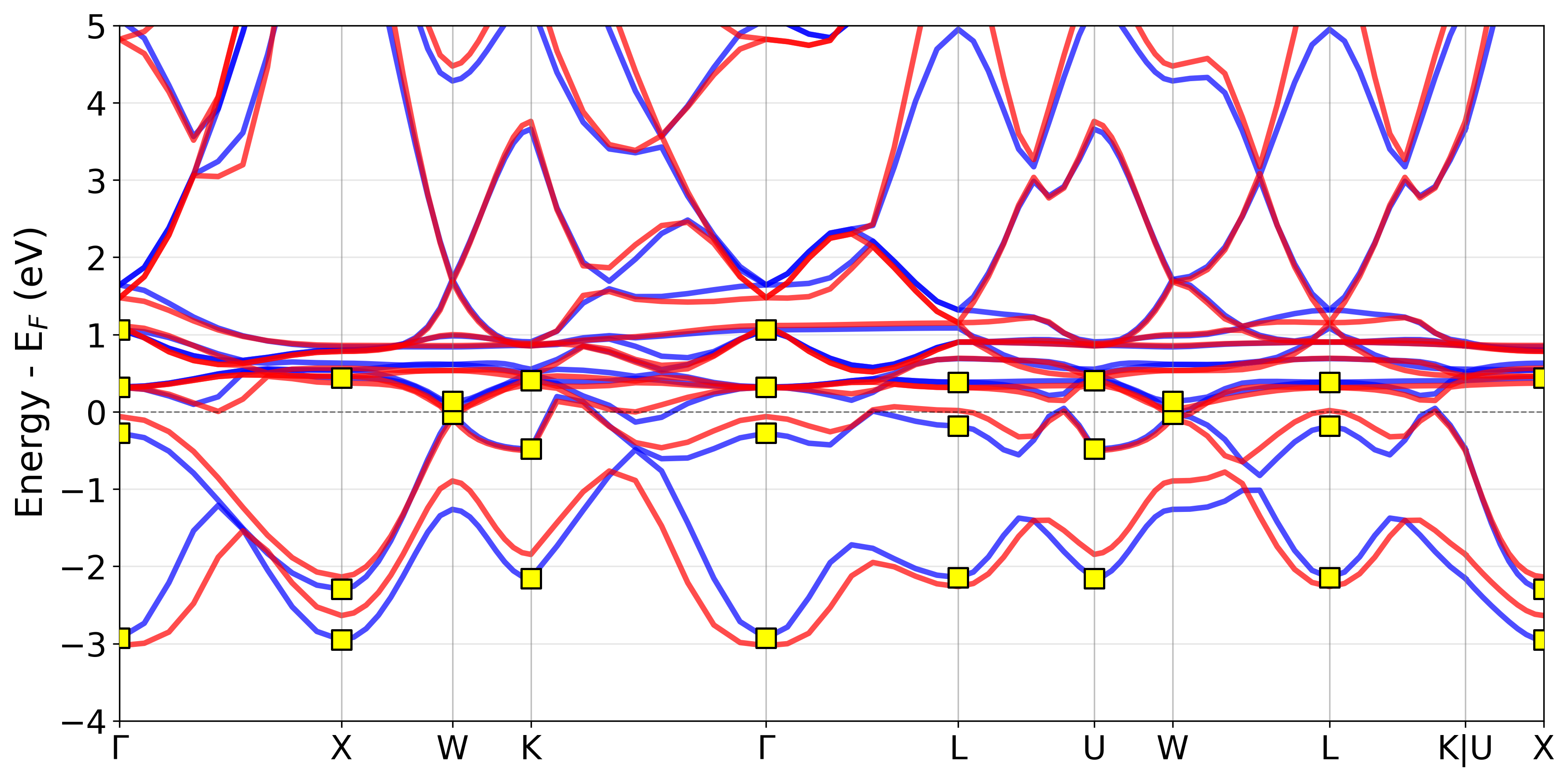}
\includegraphics[scale=0.25]{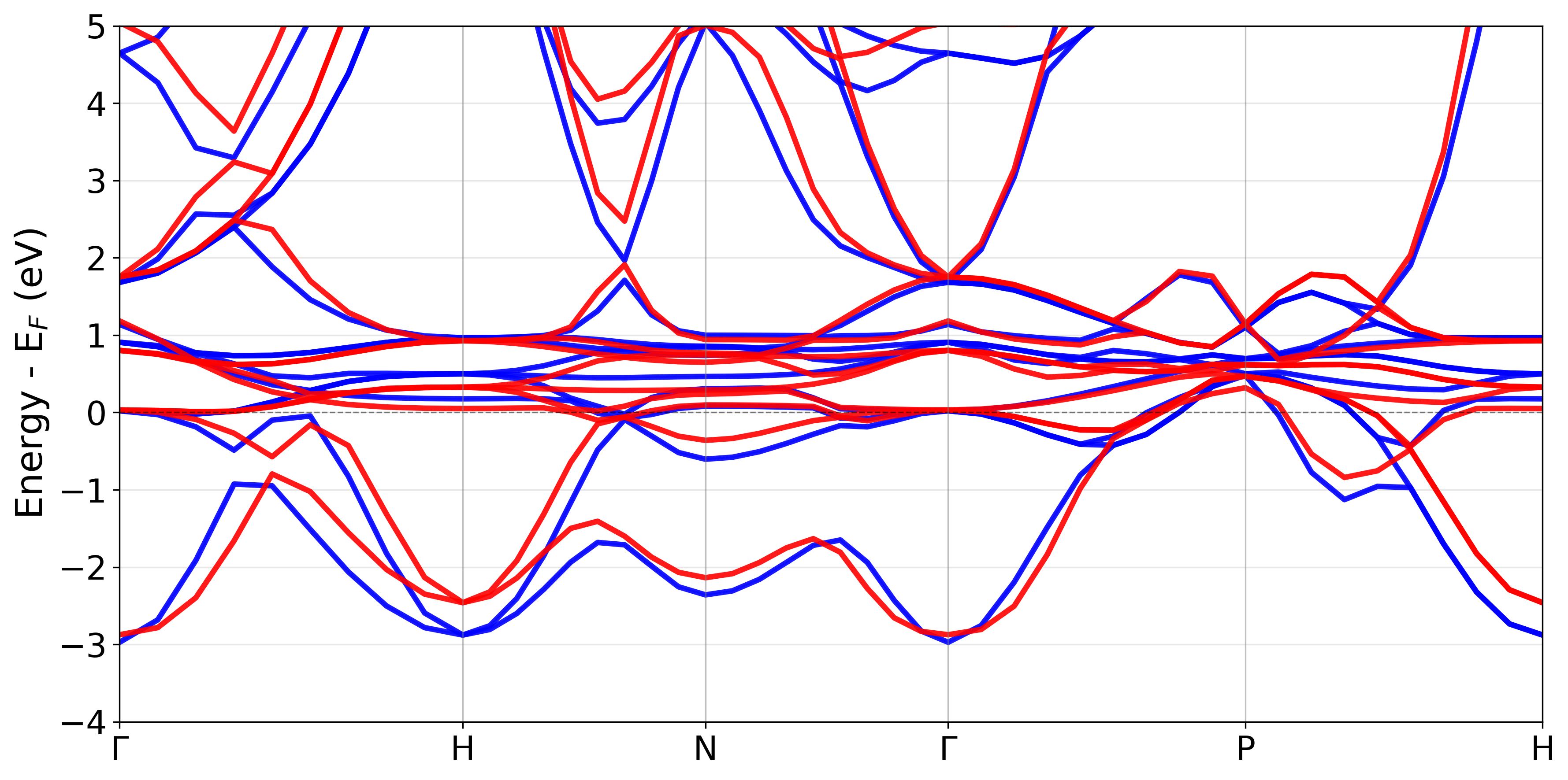}
\caption{\label{fig:bs_pbe} DFTB-PBE band structure predictions for fcc (left) and bcc cerium (right). Blue lines correspond to results from DFT and red to DFTB$+$. Yellow squares in the fcc result correspond to energies used in the DSKO decoding mask.}
\end{figure}

We first compare cerium DFTB$+$ band structure optimization results using PBE (labeled DFTB-PBE) for both $\alpha$-Ce as well as $\delta$-Ce (body centered cubic; bcc) in order to test the transferability of our model (Fig.~\ref{fig:bs_pbe}). DFTB-PBE yields similar overall band dispersion and topology for the $\alpha$ phase compared to DFT, particularly near the Fermi energy where the bands are relatively flat and non-dispersive. We observe some small disagreements for low-lying valence bands where the results from DFTB-PBE is slightly higher in energy at the $X$, $W$, $K$, and $U$ points and slightly lower at the $\Gamma$ and $L$ points. \red{The apparent degeneracies in DFT along the $\Gamma-X$ and $K-\Gamma$ lines are likely due to accidental crossings, where DFTB-PBE has shifted these bands without implying a loss of crystal symmetry.} However, we observe strong agreement for high energy conduction bands that were not included in the DSKO decoding mask. 

We also observe tight alignment for the $\delta$-Ce band dispersion and topology, which was not included in our optimization. DFTB-PBE compares well to results from DFT for states near the Fermi energy, which are slightly more flat (localized) compared to the $\alpha$ phase. We note that the band separation from DFTB-PBE at the $H$ point is slightly compressed, though this is resolved with the band separation at the $\Gamma$, $N$, and $P$ points. We also observe overall agreement for the band topologies and dispersion for the high energy conduction bands. These results are indicative of the transferability and correct GGA-level physics of DFTB-PBE for cerium allotropes, where electron localization in the $\alpha$ and $\delta$ phases is relatively similar according to this level of DFT.

\begin{figure}[!htp]
\centering
\includegraphics[scale=0.6]{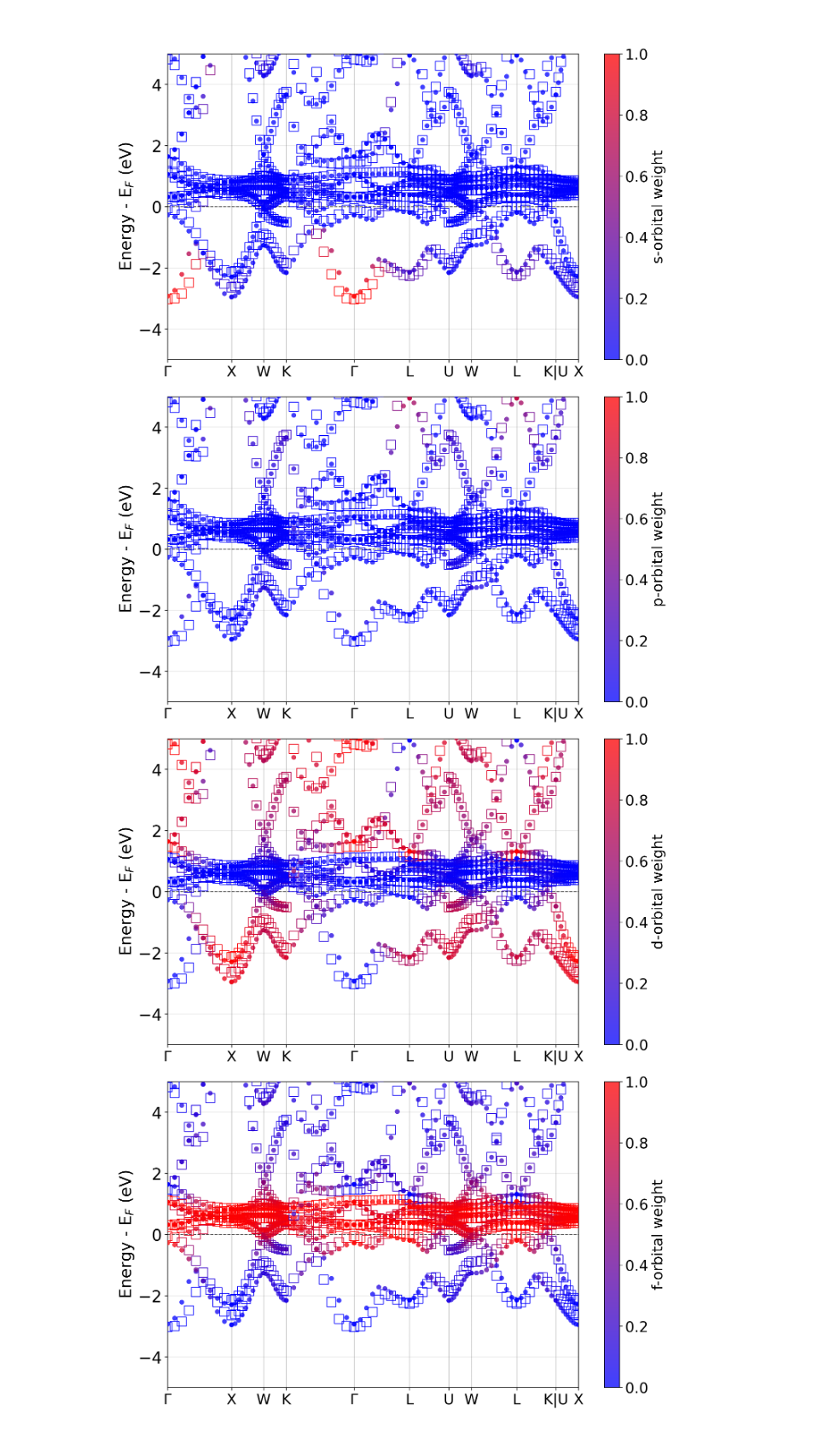}
\caption{\label{fig:orb_projections_pbe} Band structure projections for $\alpha$-Ce for s, p, d, and f-orbitals, starting from top to bottom from our DFTB-PBE model. Closed circles correspond to results from DFT and open squares to DFTB$+$. Red indicates larger orbital projection values, blue is smaller. }
\end{figure}

We have computed the orbital projections of the band structures for $\alpha$-Ce in order to validate the orbital character of the bands determined by DFTB-PBE (Fig.~\ref{fig:orb_projections_pbe}). Overall, we observe substantial overlap with DFT for all orbital channels. The s-orbital character is primarily confined to low lying valence bands around the $\Gamma$-point. The p-orbital character is largely absent from the energy window presented here, which is expected given that the Ce 6p orbitals are unoccupied and that its DFTB onsite energy was restricted to high-values as a result. The d-orbital character is primarily confined to lower energy valence bands and higher energy conduction bands, which is well represented by DFTB-PBE. Finally, DFTB-PBE correctly yields high f-orbital character around the Fermi energy, which is needed to accurately capture cerium f-electron physics. 

\begin{figure}[!htp]
\centering
\includegraphics[scale=0.25]{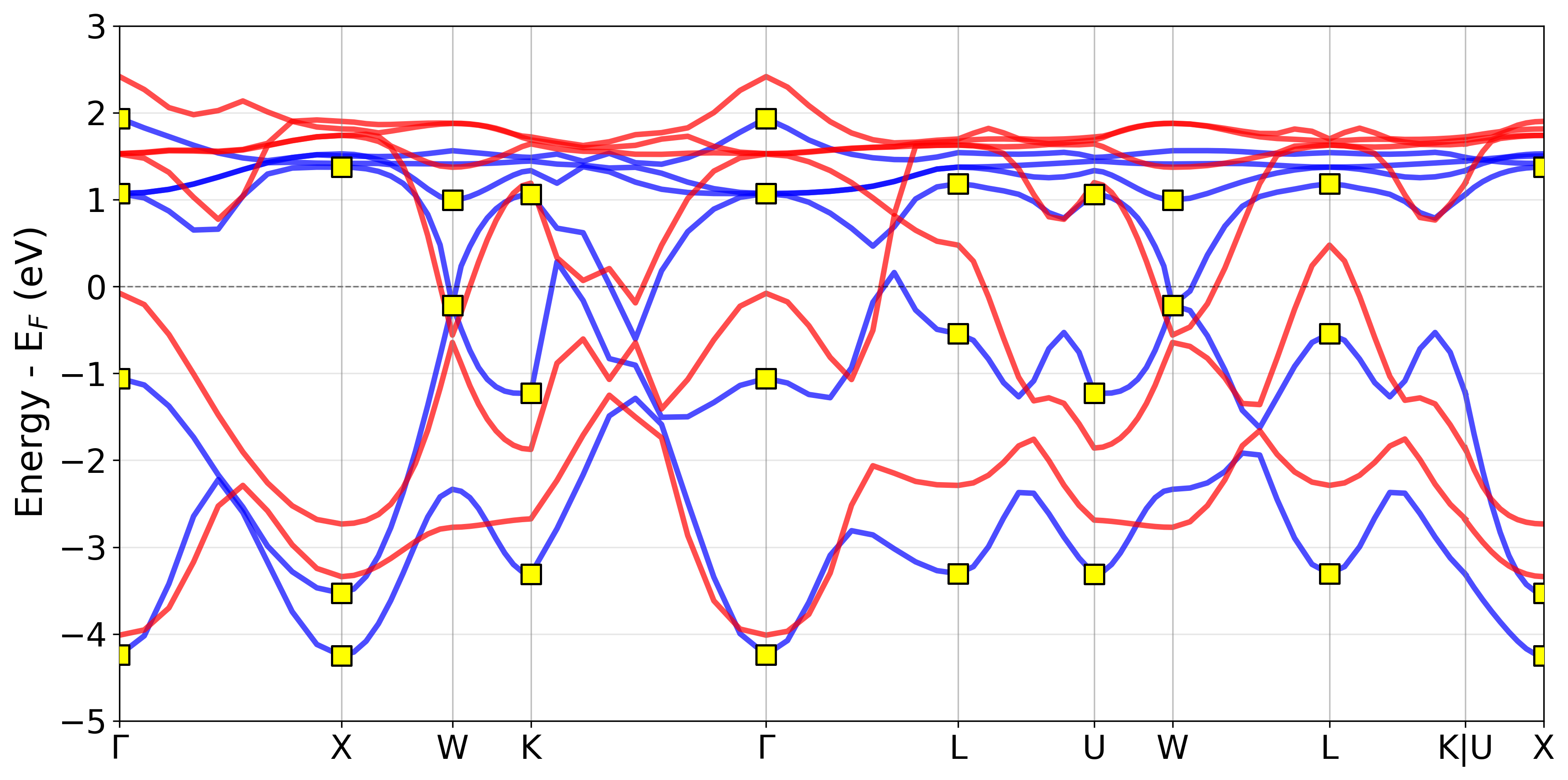}
\includegraphics[scale=0.25]{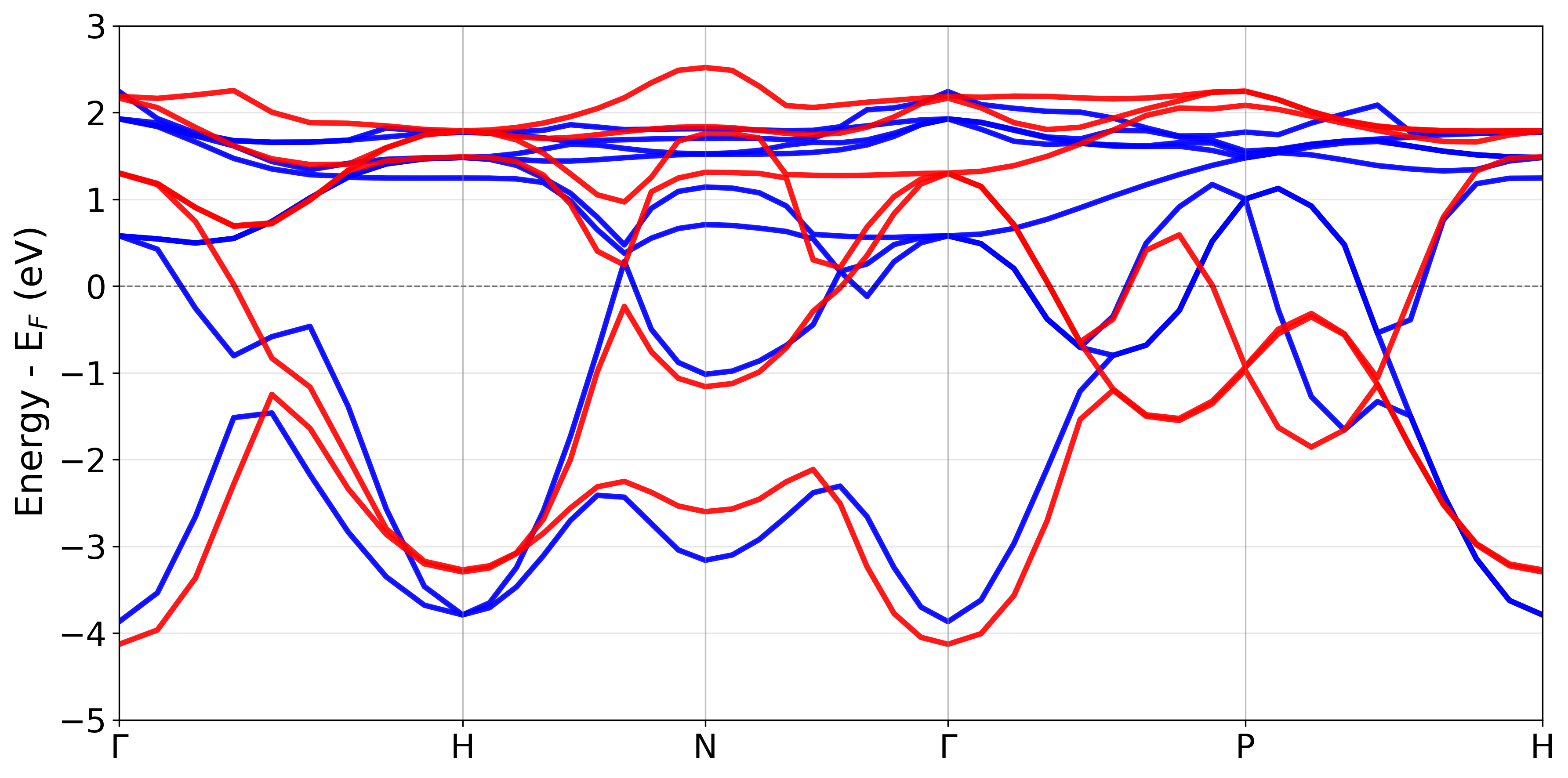}
\caption{\label{fig:bs_pbe0} DFTB-PBE0 band structure predictions for fcc (left) and bcc cerium (right). Again, blue lines correspond to results from DFT, red to DFTB$+$, and yellow squares to the DSKO decoding mask. Higher energy conduction bands are removed for the sake of clarity.}
\end{figure}

\begin{figure}[!htp]
\centering
\includegraphics[scale=0.6]{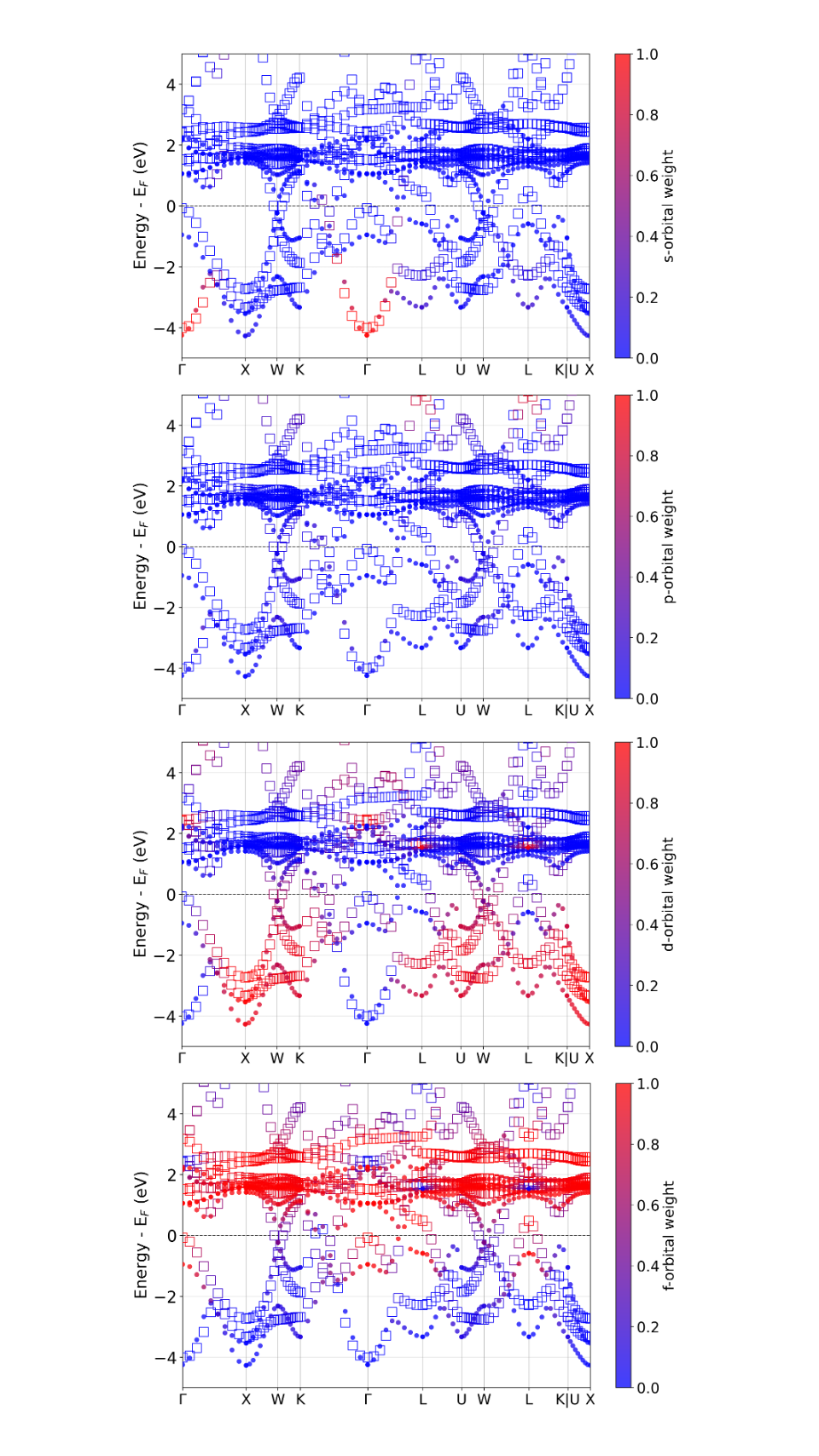}
\caption{\label{fig:orb_projections_pbe0} Band structure projections for $\alpha$-Ce for s, p, d, and f-orbitals, starting from top to bottom from our DFTB-PBE0 model. Closed circles correspond to results from DFT and open squares to DFTB$+$. Red indicates increased orbital projection values. }
\end{figure}

We now discuss DFTB$+$ band structure results from PBE0 (labeled DFTB-PBE0) for the $\alpha$ and $\delta$-Ce phases (Fig.~\ref{fig:bs_pbe0}) as well the orbital projections for the $\alpha$ phase (Fig.~\ref{fig:orb_projections_pbe0}). We note that the band separation for both DFT and DFTB-PBE0 is somewhat larger than those from PBE likely due to the increased repulsions between orbitals. The overall band topology from DFTB-PBE0 for $\alpha$-Ce is reasonable, with some deviation from DFT in terms of band dispersion and energy at specific k-points. In particular, the lowest lying valence band is disproportionately flat and high in energy at the $X$, $L$, and $U$ points. These regions of the bands largely have d-orbital character, indicative that these are too localized. In contrast, the highly dispersive valence band at the $\Gamma$ point as well as the conduction bands directly above the Fermi energy largely have f-orbital character, revealing that these electrons are too mobile and not localized sufficiently. Simultaneously, there is closer agreement with DFT for the near band crossing at the $W$ point, which has mixed d and f-character.  Our results for $\delta$-Ce yield slightly improved agreement, though high energies persist at the $H$, $N$, and $\Gamma$ points. In addition, DFTB-PBE0 underpredicts the energy of the band crossover at the $P$ point and yields fewer Fermi energy crossings in the $P- H$ direction. 

These discrepancies are likely due in part to the lack of orbital resolved SCC charge transfer in hybrid calculations with the DFTB$+$ code. The determination of a single Hubbard $U$ parameter yields charge transfer than is too hard (i.e., $U$ is too large) for the d-orbitals, resulting in excessive localization, whereas the value is too soft (i.e., $U$ is too small) for the f-orbitals, yielding high delocalization. This is confirmed by the skprogs computed d-orbital Hubbard $U$ value of 0.22 au and f-orbital value of 0.44 au, compared to the optimized value of 0.36 au used in this work.

\begin{figure}[!htp]
\centering
\includegraphics[scale=0.25]{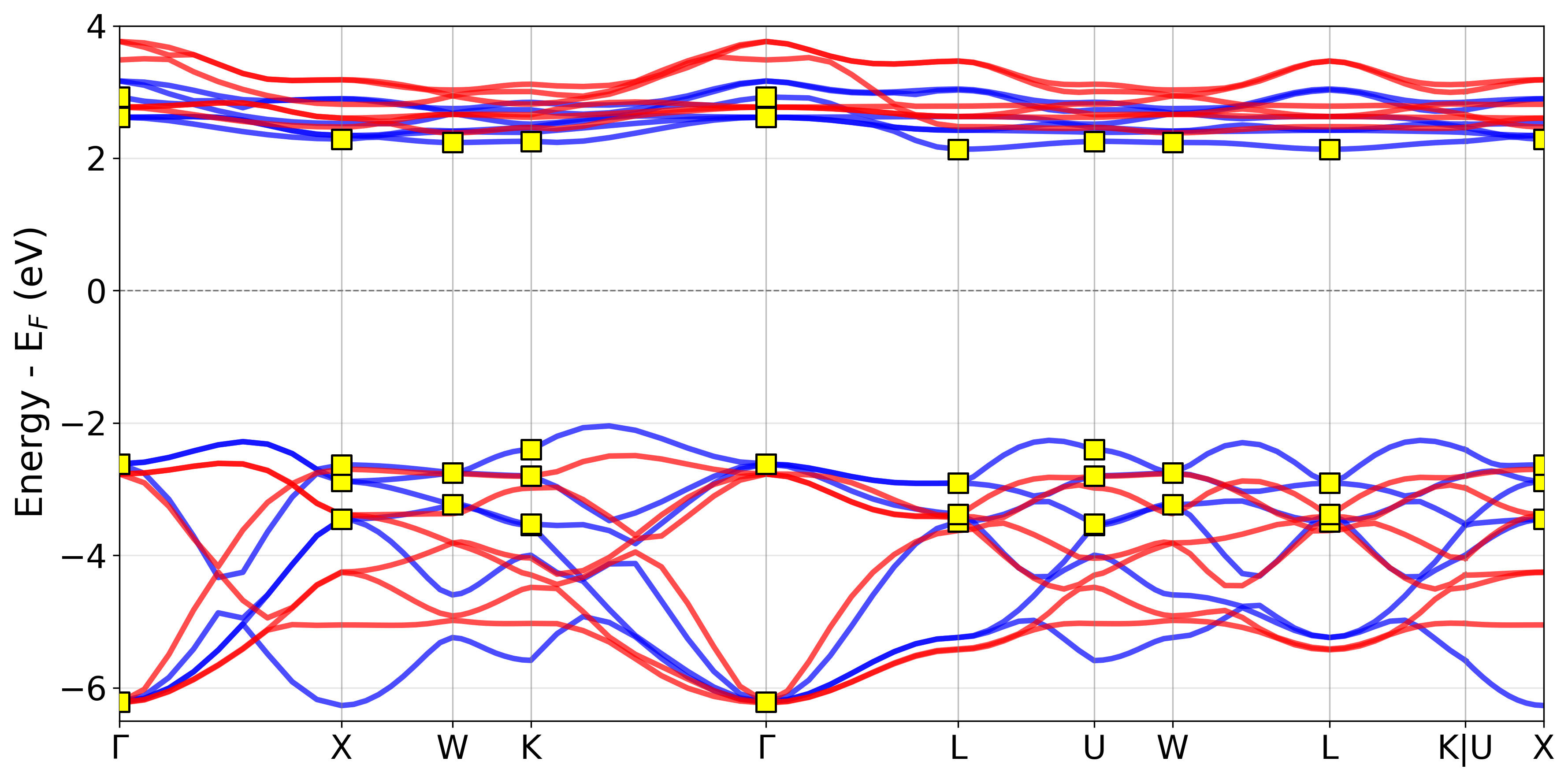}
\captionsetup{labelfont={color=black}} \caption{\label{fig:ceo2_pbe0} \red{DFTB-PBE0 band structure predictions for CeO$_2$. Again, blue lines correspond to results from DFT, red to DFTB$+$, and yellow squares to the DSKO decoding mask, with higher energy conduction bands removed for the sake of clarity. For each result, the Fermi energy is set to the mid-point between the $\Gamma$-point valence band maximum and conduction band minimum.} }
\end{figure}

\red{In order to further illustrate the utility of DFTB Slater-Koster optimization using hybrid functionals, we have determined optimal confining potentials for CeO$_2$ from PBE0  (Fig.~\ref{fig:ceo2_pbe0}). Here, we have used DSKO to optimize the ground-state fluorite-type fcc structure only using the particle swarm method with 46 particles for 43 iterations. (See Table~S6 in the Supplementary Information for decoding mask details.). We find that the topology of the band structure from DFTB matches that from DFT reasonably well, with a close match with DFT for the occupied $\Gamma$-point bands. We also find that DFTB is able to reproduce the localization or flatness of conduction bands as well as the larger dispersion for the bands at $\Gamma$ and $L$ between $3-4$~eV. The valence bands from DFTB are somewhat too localized overall, where we see a relatively flat band along most Brillouin zone path steps at approximately $-5$~eV. We also observe lower dispersive features for the DFTB valence bands between $-2$ and $-3$~eV, those these match the DFT result slightly more closely. Orbital projections indicate that the valence bands are largely due to the oxygen p-orbitals and the conduction bands to cerium f-orbitals for both DFT and DFTB (see Supplementary Information for more details). It is possible that future implementation in DFTB$+$ of orbital resolved SCC charge transfer for hybrid functional calculations will improve prediction of these properties.} 

\red{Overall, we find that our DFTB parameterization is able to match the DFT indirect band gap reasonably closely, with a value of 4.87~eV, compared to the DFT result of 4.18~eV. PBE0 is known to overestimate the CeO$_2$ band gap relative to the experimental result of 3.3~eV.\cite{wuilloud_ceo2_gap, Shi2016_PBE0_CeO2_gap} DFTB places the valence band maximum (VBM) at the k-point $\left(0.21, 0.21, 0.42\right)$ and the conduction band minimum (CBM) at the $W$ special k-point at $\left(0.50, 0.25, 0.75\right)$, whereas DFT places the VBM at $\left(0.25, 0.25, 0.50\right)$ and the CBM at the $L$ special k-point at $\left(0.50, 0.50, 0.50\right)$. However, calculation of the DFTB band gap relative to the $L$-point yields a value that is only $\sim0.09$~eV higher due to the flatness of the conduction bands.}

Next, we have completed our DFTB model development effort by determining the $E_\textrm{Rep}$ from the Chebyshev Interaction Model for Efficient Simulation (ChIMES).\cite{Goldman_SEQM_2023} \red{Our goal here is to determine possible advantages in $E_\textrm{Rep}$ optimization such as small training sets that yield a high degree of transferability by first having determined highly accurate DFTB electronic structure parameters.} Hence, for the remainder of the work, we choose to focus on \red{cerium metal} DFTB-PBE only due to its more accurate band structure prediction relative to the underlying DFT method. ChIMES is an ideal approach in this case as it allows us to systematically increase the complexity of  $E_\textrm{Rep}$ through inclusion of greater than 2-body interactions.  Briefly, the ChIMES potentials are determined by considering the DFT total energy as a sum of atomic cluster interactions with increasing size:

\begin{equation}
E_{\mathrm{DFT}} =
\sum^{n_\mathrm{a}}_{i_1}{}^1 E_{i_1} + \sum^{n_\mathrm{a}}_{i_1>i_2}{}^2 E_{i_1 i_2} + \sum^{n_\mathrm{a}}_{i_1>i_2>i_3}{}^3 E_{i_1 i_2 i_3} + \sum^{n_\mathrm{a}}_{i_1>i_2>i_3>i_4}{}^4 E_{i_1 i_2 i_3 i_4}  + \dots + \sum^{n_\mathrm{a}}_{i_1>i_2\dots\ i_{n{_\mathrm{B}-1}}>i_{n{_\mathrm{B}}}}{}^{n_\mathrm{B}}\!E_{i_1 i_2 \dots i_n{_\mathrm{B}}}.
\label{eq:genchimes}
\end{equation}

\noindent Here, the one-body terms, ${}^1\!E_{i_1}$, correspond to the atomic energy constants, the two-body (2B) terms, ${}^2\!E_{i_1 i_2}$, to all pair-wise energies, the three-body terms (3B), ${}^3\!E_{i_1 i_2 i_3}$, to all triplet energies, etc., all the way up to some predetermined maximum cluster size, $n_\mathrm{B}$.  In general, ChIMES models are determined with up to four-body (4B) interactions. Each of these interactions is determined by a linear combination of 2B or greater Chebyshev polynomials, where many-body polynomials are created through the tensorial product of their constituent 2B pairwise polynomial interactions. The sum is performed over all $n_\mathrm{a}$ atoms in the system and includes all clusters within defined radial cutoffs. More details can be found in Refs.~\citenum{Goldman_TiH2, Goldman_SEQM_2023, Pham_DFTB_JPCL, chimes_carbon_2.0, chimes_hierarchical_2026} as well as the Supporting Information, where we discuss the optimization process and show the root mean square errors for the cerium repulsive energies. 

For this work, our $E_\textrm{Rep}$ training set was determined from 32 atom $NVT$ simulations run at 600~K with VASP using the PBE functional, a $2 \times 2 \times 2$ k-point mesh, and Nos\'e-Hoover thermostats to control the temperature.\cite{Nose84, Hoover85} Simulations were run with a timestep of 2.0~fs at the optimized density as well as with the supercell lattice vectors scaled by a factor of 0.98 and 1.02 in order to sample compressed and expanded configurations. Each trajectory was run for 6-8~ps, after which configurations were sampled at a regular interval of 300~fs. This resulted in a training set consisting of a total of 143 configurations, or 13,871 data points, consisting of the atomic forces and system total energies. \red{Our effort is thus substantially smaller than previous $E_\mathrm{Rep}$ machine-learning approaches, which required larger and more varied training sets (i.e., approximately between 7,000 to 500,000 configurations).\cite{Margraf_GPR_2020, Yang_MLTB_2025, Tkatchenko_2026}} 

We have then created a DFTB/ChIMES model with two-body interactions only (order 12; labeled DFTB/ChIMES (2B)), a second with two and three-body interactions (orders 12 and 8, respectively; labeled DFTB/ChIMES(3B)), and a third with up to four-body interactions (orders 12, 8, and 4, respectively; labeled DFTB/ChIMES (4B)). These polynomial orders are consistent with previous choices for ChIMES repulsive energies for condensed matter.\cite{Goldman_SEQM_2023} This set of models allows us to asses the effect of many-body repulsive energies on the accuracy of our DFTB models for dynamic and structural properties. 

\begin{table}[!htp]
  \footnotesize
  \caption{Nearest neighbor distances (NN) in \AA\  and ground state energies relative to $\alpha$-Ce ($\Delta E_\alpha$) in meV/atom for the cerium polymorphs considered in this work.}
  \label{tab:ce_polymorphs}
\begin{tabular}{rl| cc| cc| cc| cc}
\hline
\multicolumn{2}{c}{} &  \multicolumn{2}{c}{$\alpha$ (fcc)} & \multicolumn{2}{c}{$\beta_1$ (dhcp)} &  \multicolumn{2}{c}{$\beta_2$ (dhcp)} &  \multicolumn{2}{c}{$\delta$ (bcc)}  \\
\hline
\hline
& & NN & $\Delta E_\alpha$ & NN & $\Delta E_\alpha$ & NN & $\Delta E_\alpha$ & NN  & $\Delta E_\alpha$  \\
DFTB/ChIMES & (2B)  & 3.29 & 0 & 3.29 & 27 & 3.26 & 56 & 3.20 & 142  \\
                         & (3B)  & 3.33 & 0 & 3.33 & 26 & 3.30 & 50 & 3.24 & 171  \\
                         & (4B)  & 3.33 & 0 & 3.33 & 26 & 3.29 & 49 & 3.24 & 187  \\
DFT                  &          & 3.34 & 0 & 3.37 & 56 & 3.27 & 87 & 3.27 & 227  \\
\hline
\end{tabular}
\end{table}

We have validated our DFTB/ChIMES models by computing the optimized properties of four chosen cerium allotropes found in the Materials Project database. These include the fcc $\alpha$ and bcc $\delta$ phases, as well as two double hexagonal close packed (dhcp) phases, one with lower symmetry (four atoms per primitive cell, labeled $\beta_1$) and a second with higher symmetry (two atoms per primitive cell, labeled $\beta_2$).  Our results indicate that all DFTB/ChIMES models consistently match the DFT-computed energetic ordering of the phases, with $\alpha$ as the ground-state, followed by $\beta_1$, $\beta_2$, and the $\delta$ phase (Table~\ref{tab:ce_polymorphs}). The energetic differences relative to the $\alpha$ phase are relatively consistent compared to DFT. In this case, all three DFTB/ChIMES models are $29-30$ meV too low for the $\beta_1$ phase and $31-38$ meV too low for $\beta_2$. These models show some differentiation with respect to the $\delta$ phase, where DFTB/ChIMES (2B) underpredicts the energy by 85 meV, DFTB/ChIMES (3B) by 56 meV, and DFTB/ChIMES (4B) by 40 meV. We also determine some difference in results for the nearest neighbor distance for each phase, where DFTB/ChIMES (3B) and (4B) tend to yield slightly improved agreement with DFT compared to DFTB/ChIMES (2B). Here, DFTB/ChIMES (3B) and (4B) yield nearest neighbor distance errors of 0.3\% for the $\alpha$ phase, 1.2\% for $\beta_1$, 0.9\% and 0.6\% respectively for $\beta_2$, and 0.9\% for the $\delta$ phase. In contrast, DFTB/ChIMES (2B) yields errors of 1.5\% for $\alpha$, 2.4\% for $\beta_1$, 0.3\% for $\beta_2$, and 2.1\% for the $\delta$ phase. Our results indicate that higher ChIMES \red{body interactions} can yield some additional accuracy, though the results from DFTB/ChIMES (2B) are reasonable.

\begin{figure}[!htp]
\centering
\includegraphics[scale=0.3]{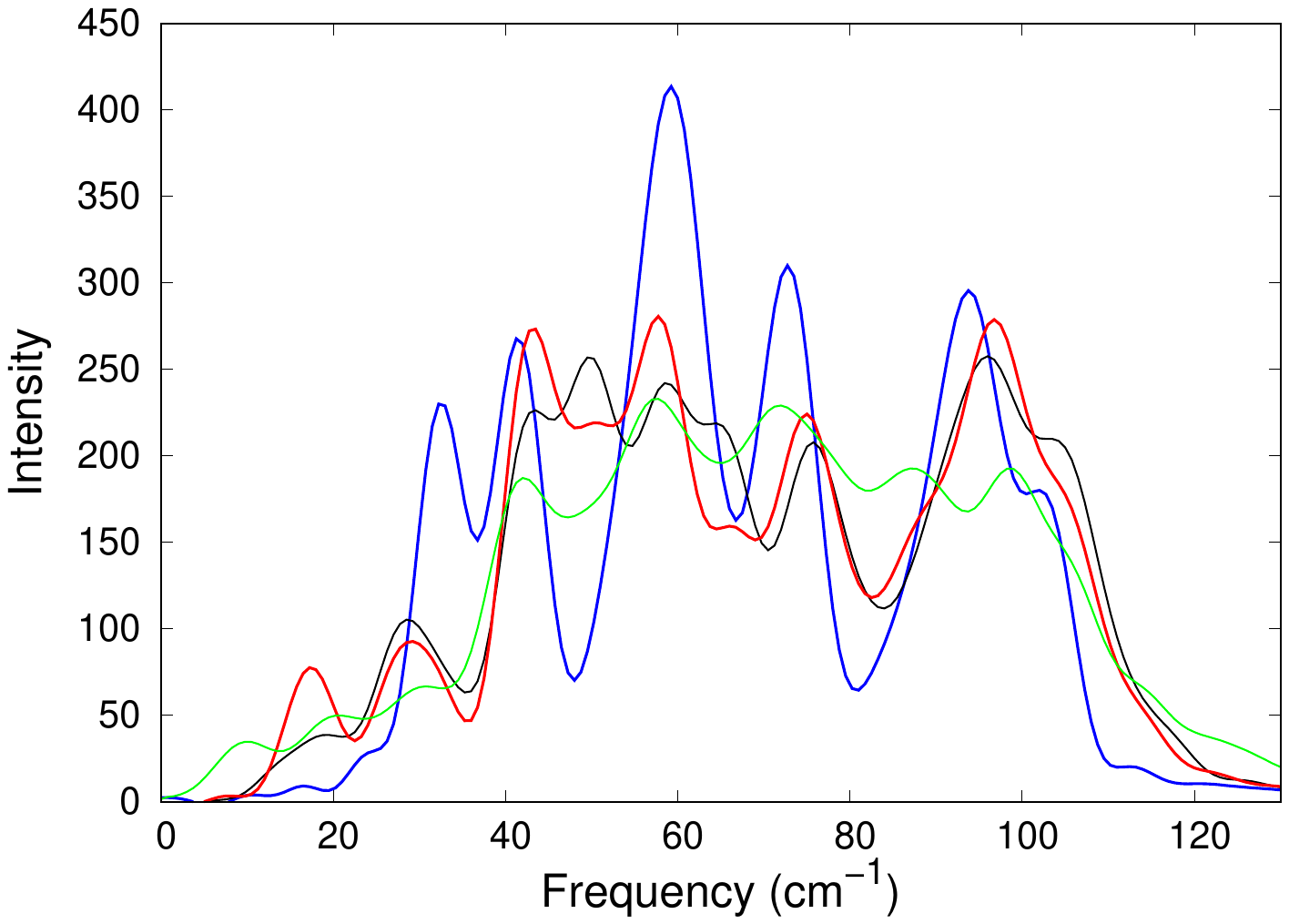}
\caption{\label{fig:pdos} Vibrational density of states computed from MD simulations with DFTB-PBE at 600~K. The blue line corresponds to results from DFT, the red line to 3-body DFTB/ChIMES, black to 4-body DFTB/ChIMES, and the green to 2-body DFTB/ChIMES. \red{Fourier transform results have been smoothed with a Welch window function.}}
\end{figure}

Finally, we assess the force output from each model through comparison of the resulting vibrational density of states (VDOS) for the $\alpha$ phase (Fig.~\ref{fig:pdos}), computed by Fourier transform of the velocity autocorrerlation function. These were determined by running MD with both DFT and DFTB/ChIMES using a 108 atom supercell at 600~K for 8--10 ps with a  $2 \times 2 \times 2$ k-point mesh and timestep of 2.0~fs. Our DFT calculations had an approximate wall-clock time of $\sim$97,000 CPU-seconds per timestep (computed as the number of CPUs utilized multiplies by the average wall clock time per step) on Intel Xeon Ice Lake Gold 2.8 GHz processors, compared to a value of $\sim$1,100 for DFTB/ChIMES, indicating an increase of around $100\times$ in computational efficiency. Calculation of the ChIMES repulsive energies added negligible time to our DFTB calculations, which were dominated by the matrix diagonalization steps. 

DFTB/ChIMES (3B) yields the most accurate results overall, with peaks matching closely with those from DFT at $\sim$41, 59, 73, and 94~cm$^{-1}$. However, our model yields peak intensities that are somewhat low at several frequencies (e.g., 59 and 73~cm$^{-1}$), as well as two small peaks that are not present in DFT (18 and 50~cm$^{-1}$). These frequencies are relatively low and thus can be poorly sampled by our training set and MD trajectory, whereas DFTB/ChIMES (3B) appears to yield higher accuracy around the vibron frequency (close to 94~cm$^{-1}$), which likely has improved sampling overall. 

DFTB/ChIMES (4B) is the next most accurate model, where we observe somewhat of a loss of features of the vibrational spectrum, with peaks decreasing in intensity and the addition of several new peaks not present in the DFT spectrum between 40 and 70~cm$^{-1}$. However, we note that this model does yield a reasonable vibron peak (as well as the small right shoulder at high frequency) and that the peak at 18~cm$^{-1}$ from the (3B) result is now absent. \red{The lower quality of the 4B VDOS result is likely due to overfitting of the ChIMES energy function.} DFTB/ChIMES (2B) yields the poorest agreement with DFT overall, with the resulting vibrational spectrum showing fewer distinct peaks or features, including a diminished and flattened vibron peak.

Overall, we observe that DFTB/ChIMES can yield models that retain a significant portion of the accuracy of the electronic states and physical properties of DFT from relatively small amounts of training data. These models can exhibit a high degree of transferability, where our models reproduce the DFT computed band structure for cerium polymorphs as well as the relative energies of solid phases with different crystal symmetries despite only being tuned to data from the fcc phase. \red{We note that our DFTB models may not be suited for more extreme temperatures and pressures, as the band structure and repulsive energies are tuned only to lower conditions.} Our DFTB/ChIMES models can also produce stable molecular dynamics trajectories and accurate vibrational densities of states with close to two orders of magnitude increase in computational efficiency. \red{Our efforts lend credence to the hypothesis that careful optimization of the DFTB band structure can streamline $E_\mathrm{Rep}$ determination and thus allow for small training sets to help determine transferable models. The training set used here is somewhat smaller than previous DFTB/ChIMES efforts as well.\cite{Goldman_TiH2, Goldman_SEQM_2023, Dettori2025} } \red{We have also shown the possibility of using DFTB with hybrid functionals to study gapped materials that might otherwise require significantly more computationally expensive DFT calculations.} Future work in this area can involve the parameterization of additional f-electron metals and ceramics, where DFT calculations can be computationally inefficient and there is a strong need for new methods that can compute both electronic and physical material properties with reduced computational effort.   

\section*{Acknowledgments}

This work was performed under the auspices of the U.S. Department of Energy by Lawrence Livermore National Laboratory under Contract DE-AC52-07NA27344. 

\section*{Data and Software Availability}

\red{Our modified DSKO code is available in the main branch of the public repository: 
\\
https://gitlab.mpcdf.mpg.de/fhi-theory/DSKO.} 

The ChIMES training code and calculator are available online.\cite{chimes_code, chimes_lsq} DFTB$+$ skf files and ChIMES repulsive energies for cerium are available for download at 
\\
https://github.com/nirgoldman/dftb\_cerium\_oxygen.git. 
\\
\\
\emph{Supporting Information:} optimized DSKO output for the DFTB$+$ Slater-Koster interaction files, \red{additional details regarding DSKO code modifications,} the ChIMES repulsive energy parameters and optimization details, \red{and CeO$_2$ band structure orbital projections.}

\newpage	

\bibliography{library}

\newpage
\begin{figure}
\begin{center}
\includegraphics[scale=1.5]{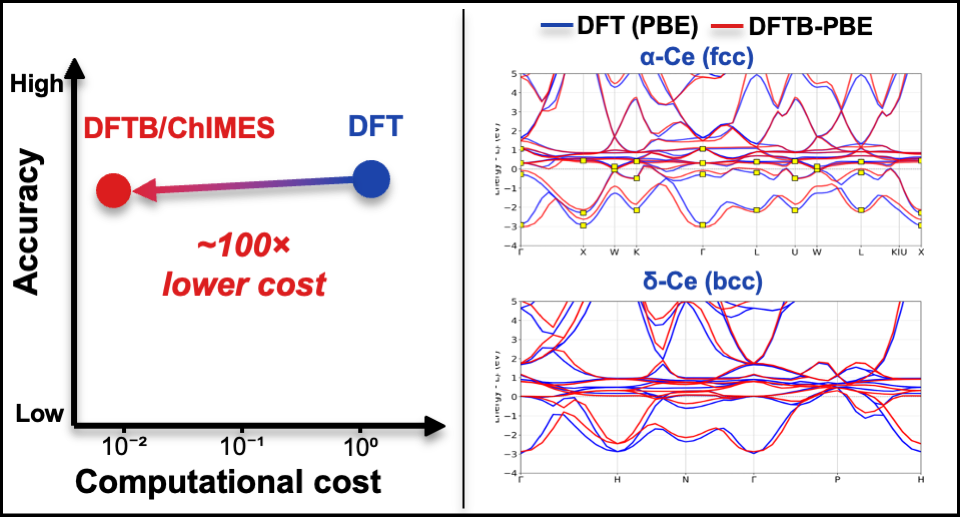}
\caption*{\textbf{TOC Graphic}}
\end{center}
\end{figure}

\end{document}


\newpage

\red{\section{Modifications to DSKO}
We have included new options in DSKO to allow for relativistic calculations for higher-Z elements (through the Zeroth-Order Regular Approximation) and the fraction of exact exchange for the hybrid functional. PBE0 calculations are performed using the truncated Coulomb matrix method with a cutoff of 20.0 au and neighbour-based screening with a threshold of $10^{-5}$. These parameters were validated against $\alpha$-Ce calculations using a cutoff of 40.0 au and a threshold of $10^{-5}$. We enforce an odd k-point mesh, as required for many hybrid functional calculations in order to avoid singularities at the $\Gamma$-point. Shell resolved SCC is also turned off, as required for hybrid functionals by the current DFTB$+$ release.}

\section{ChIMES Methodology}
An in-depth review of ChIMES can be found in Ref.~\citenum{Goldman_SEQM_2023}. Briefly, in the ChIMES formalism, we represent each of the terms in our n-body energy expansion as a linear combination of Chebyshev polynomials. Chebyshev polynomials of the first kind of order $m$ are defined by the expression $T_m \left(\mathrm{cos} \, \theta \right) = \mathrm{cos}\left(m \theta\right)$, more commonly written as $T_m(x)$, where $x = \mathrm{cos}\, \theta$ and thus exists over the range $\left[-1,1\right]$. The derivatives of Chebyshev polynomials of the first kind are related to Chebyshev polynomials of the second kind $U_m(x)$ by the expression $\mathrm{d}T_m/\mathrm{d}x = m U_{m-1}$, where $U_m\left( \mathrm{cos} \, \theta \right) = \mathrm{sin} \left[ \left(n+1\right) \theta \right]/\mathrm{sin} \, \theta$.

We write the two-body (2B) energy term as the following expression:

\begin{equation}
        {}^2\!E_{i_1 i_2} = f_{\mathrm{p}}\left(r_{i_1 i_2}\right) + f^{e_{i_1} e_{i_2}}_{\mathrm{c}}\left(r_{i_1 i_2}\right)
             \sum^{\mathcal{O}_{2}}_{m=1} C_m^{e_{i_1} e_{i_2}} T_m (s^{e_{i_1} e_{i_2}}_{i_1 i_2})
        \label{eqn:FF2B}
\end{equation}

\noindent $C_{m}^{e_{i_1} e_{i_2}}$ is an optimized coefficient for the interaction between atom types $e_{i_1}$ and $e_{i_2}$, taken from the set of all possible element types, $\{\boldsymbol{e}\}$. All $C_{m}^{e_{i_1} e_{i_2}}$  are permutationally invariant. $T_{m}\left(s^{e_{i_1} e_{i_2}}_{i_1 i_2}\right)$ represents a Chebyshev polynomial of order $m$, and $s^{e_{i_1} e_{i_2}}_{i_1 i_2}$ is the pair distance transformed to occur over the interval $[-1,1]$ using a Morse-like function\cite{WHBB-2009,WHBB-2011}. For that coordinate transform, $s^{e_{i_1} e_{i_2}}_{i_1 i_2} \propto \mathrm{exp}\left(-r_{i_1 i_2}/\lambda_{e_1 e_2}\right)$ and $\lambda_{e_1 e_2}$ is an element-pair distance scaling constant, frequently set to the first peak in a radial distribution function. Further details are discussed in Ref.~\citenum{Lindsey17}. The term $f^{e_{i_1} e_{i_2}}_\mathrm{c}(r_{i_1 i_2})$ is a Tersoff cutoff function\cite{Tersoff88} which smoothly varies to zero up to a predefined maximum cutoff distance. In order to prevent sampling of $r_{i_1 i_2}$ distances below those sampled in our training set, we introduce use of a smooth repulsive penalty function $f_\mathrm{p} (r_{i_1 i_2})$ that is non-zero for distances close to the inner cutoff of the Chebyshev polynomials.

Many-body (e.g., greater than two-body) orthogonal polynomials can be created by defining a cluster of size $n$ and taking the product of the Chebyshev polynomials derived from the constituent ${n \choose 2}$ unique pairs. For example, the three-body polynomials will be products of ${3 \choose 2} = 3$ two-body polynomials. We thus write the ChIMES three-body (3B) energy as the following:

\begin{equation} 
{}^3 E_{i_1 i_2 i_3} = f^{e_{i_1} e_{i_2}}_{\mathrm{c}}\left(r_{i_1 i_2}\right) f^{e_{i_1} e_{i_3}}_{\mathrm{c}}\left(r_{i_1 i_3}\right) f^{e_{i_2} e_{i_3}}_{\mathrm{c}}\left(r_{i_2 i_3}\right) \sum^{\mathcal{O}_{3}}_{m=0} \sum^{\mathcal{O}_{3}}_{p=0} {\sum^{\mathcal{O}_{3}}_{q=0}}^\prime C^{e_{i_1} e_{i_2} e_{i_3}}_{mpq} T_{m}\left(s^{e_{i_1} e_{i_2}}_{i_1 i_2}\right) T_{p}\left(s^{e_{i_1} e_{i_3}}_{i_1 i_3}\right) T_{q}\left(s^{e_{i_2} e_{i_3}}_{i_2 i_3}\right).
        \label{eqn:FF3B}
\end{equation}

\noindent We thus compute a triple sum for the product of the $i_1 i_2$, $i_1 i_3$, and $i_2 i_3$ pair-wise polynomials. These are computed up to a predefined order ($\mathcal{O}_{3}$) for each three-body polynomial and then multiplied by a single coefficient, $C^{e_{i_1} e_{i_2} e_{i_3}}_{mpq}$, that is permutationally invariant for each set of polynomial orders and atom types. The primed sum in Equation~\ref{eqn:FF3B} indicates that only terms for which two or more of the $m,p,q$ polynomial powers are greater than zero are included in order to guarantee that all atoms in the cluster interact through Chebyshev polynomials.  
The expression for ${}^3\!E_{i_1 i_2 i_3}$ also contains the $f_{\mathrm{c}}$ smoothly varying cutoff functions for each constituent pair distance. The penalty function $f_p$ is not included here.

Higher bodied terms are included in ChIMES in a similar fashion. For example, four-body (4B) terms are often included in ChIMES optimizations\cite{Pham_HN3_2021}, where ${}^4\!E_{i_1 i_2 i_3 i_4}$ is now determined from the sum over the product of the ${4 \choose 2} = 6$ constituent pair-wise polynomials multiplied by a single permutationally invariant coefficient. In the determination of permutational invariance for an arbitrary number of bodies, it is important to realize that the atom indices and atom types are permuted, which then implies a corresponding permutation of the bond distance indices in $C$.   

Similar to other machine learning atomic interaction potentials, ChIMES uses the method of force matching\cite{Ercolessi94} to determine the interaction parameters.
In force matching a training set of quantum simulations is generated with differing configurations.  For each configuration in the training set, the quantum mechanical
energy, atomic force, and stress (for condensed phase configurations) are calculated. In addition, weights can be used when matching to forces, energies, and stresses, due to the differing physical units and number of parameters per configuration.  In this work, we use a weight of $5/n_{a}$ for the energies ($n_a$ is equal to the number of atoms in a give configuration), and a value of one for the forces. Stress tensor components have been omitted from these optimizations, though we include them in our discussion for the sake of clarity. Once weights are determined, an objective function for optimization may be defined as follows:

\begin{footnotesize}
\begin{equation}
        F_{\mathrm{obj}}  =  \frac{1}{N_d} \sum_{\tau=1}^M  \left(\sum_{i=1}^{N_\tau} \sum_{\alpha=1}^{3} \left(w_{\mathrm{F}}\Delta \mathrm{F}_{\tau_{\alpha_i}} \right)^{2} + \sum_{\alpha=1}^3 \sum_{\beta \le \alpha}\left(w_{\sigma}\Delta\sigma_{\tau_{\alpha\beta}}\right)^{2} + \left(w_{\mathrm{E}} \Delta E_{\tau}\right)^{2}
\right)
.
\label{eqn:rmse}
\end{equation}
\end{footnotesize}

\noindent Here, $\tau$ corresponds to a specific training set configuration, $i$ is the atomic index, and $\alpha$ and $\beta$ are the cartesian directions. 
$M$ is the total number of configurations in the training set and $N_d = 3\sum_\tau N_\tau + 7 M$ is the total number of data entries (6 stress tensor components and one energy value per configuration).  In addition, $\Delta F_{\tau_{\alpha_i}} = F^{\mathrm{ChIMES}}_{\tau_{\alpha_i}} - F^{\mathrm{DFT}}_{\tau_{\alpha_i}}$, $\Delta\sigma_{\tau_{\alpha\beta}} = \sigma^{\mathrm{ChIMES}}_{\tau_{\alpha\beta}} - \sigma^{\mathrm{DFT}}_{\tau_{\alpha\beta}}$, 
and $\Delta E_{\tau} = E^{\mathrm{ChIMES}}_{\tau} - E^{\mathrm{DFT}}_{\tau}$.  The value $w_{\mathrm{F}}$ is the weight for forces, $w_{\sigma}$ for stresses, and $w_{\mathrm{E}}$ for energies. 

Optimal ChIMES parameters are determined by solving 
\begin{equation}
\frac{\partial F_{\mathrm{obj}}}{\partial C_I} = 0,
\label{eqn:optF}
\end{equation}
for all $I$, where $I$ is a combined index of the permutationally unique coefficient. 
Optimizing $F$ is equivalent to solving the overdetermined matrix equation 
\begin{equation}
\boldsymbol{w} \boldsymbol{A} \boldsymbol{C} = \boldsymbol{w} \boldsymbol{B}.
\label{eqn:linear}
\end{equation}
The matrix $\boldsymbol{A}$ corresponds to the derivatives of the ChIMES energy, stress, or force expression with respect to the fitting coefficients.  The column vectors $\boldsymbol{C}$ and $\boldsymbol{B}$ correspond to the ChIMES coefficients to be optimized and the numerical values for the training data, respectively. The diagonal matrix $\boldsymbol{w}$ is comprised of the weights to be applied to the
elements of $\boldsymbol{B}$ and rows of $\boldsymbol{A}$. Solution to this linear least-squares problem can performed using a number of different optimization algorithms, discussed in Ref.~\citenum{Goldman_SEQM_2023}.

\section{ChIMES parameters and fitting results}
For this each of the ChIMES repulsive energies discussed in this work, we have used a minimum radial cutoff of 2.35 \AA\, where 2.40 \AA\ was the minimum cerium nearest neighbor distance sampled in our training set.  We have set the ChIMES. maximum cutoff to a value of 4.35 \AA\, which ensures inclusion of second-nearest neighbor interactions for the molecular dynamics simulations discussed in the manuscript. Use of a cutoff that included nearest-neighbor interactions only (e.g., 3.40 \AA) resulted in similar results for the energetic ordering of the cerium polymorphs but worse agreement with DFT for the vibrational density of states. Optimizations were performed with the LASSO/LARS algorithm, using a regularization parameter value of $1 \times 10^{-4}$, and included energy and force data, only.

\begin{table}[!htp]
\caption{Root mean squared (RMS) errors from our ChIMES repulsive energy optimizations.}
\label{tab:rms}
\begin{tabular}{ccc}
\hline
ChIMES $E_\textrm{Rep}$ bodiedness  & Force (eV/\AA) & Energy (eV/atom)  \\
\hline
\hline
2B & 0.21 & 0.06 \\
3B & 0.14 &  0.05 \\
4B & 0.14 &  0.05  \\
\hline
\end{tabular}
\end{table}

\section{DSKO optimized DFTB-PBE parameters}
\begin{table}[!htp]
\caption{Woods-Saxon confinement parameters and Hubbard $U$ values for Ce.}
\label{tab:ce_params_pbe}
\begin{tabular}{ccccc}
\hline
Shell & $r_0$ & $a$ & $W$ & Hubbard $U$ (eV) \\
\hline
\hline
4f & 4.1068 & 13.5678 & 1.2146 & 0.4670 \\
5d & 4.5135 & 60.0000 & 6.0000 & 0.2247 \\
6s & 4.0472 & 60.0000 & 1.0000 & 0.1640 \\
6p & 5.9384 & 59.0451 & 1.3237 & 0.2247 \\
\hline
Density & 23.8488 & 5.0000 & 18.2151 & --- \\
\hline
\end{tabular}
\end{table}

\begin{table}[!htp]
\caption{PBE spin constant matrix for Ce (in eV).}
\label{tab:ce_spin_constants_pbe}
\begin{tabular}{c|cccc}
\hline
 & 4f & 5d & 6s & 6p \\
\hline
4f & -0.01235 & -0.00245 & -0.00078 & -0.00245 \\
5d & -0.00245 & -0.00902 & -0.00582 & -0.00902 \\
6s & -0.00078 & -0.00582 & -0.01009 & -0.00582 \\
6p & -0.00245 & -0.00902 & -0.00582 & -0.00902 \\
\hline
\end{tabular}
\end{table}

\newpage 
\section{DSKO optimized DFTB-PBE0 parameters}

\begin{table}[!htp]
\caption{Confinement parameters and Hubbard $U$ values for Ce. Shell-resolved SCC is not available for hybrid functionals in the DFTB$+$ code,\cite{DFTB+_2025} hence the equal values for all $U$.}
\label{tab:ce_params_pbe0}
\begin{tabular}{ccccc}
\hline
Shell & $r_0$ & $a$ & $W$ & Hubbard $U$ (eV) \\
\hline
\hline
4f & 11.6330 & 25.3199 & 2.3849 & 0.3590 \\
5d & 5.8006 & 42.1194 & 4.1573 & 0.3590 \\
6s & 5.2377 & 38.0321 & 2.3187 & 0.3590 \\
6p & 3.9278 & 31.0841 & 1.1961 & 0.3590 \\
\hline
Density & 20.1927 & 34.5318 & 36.2038 & --- \\
\hline
\end{tabular}
\end{table}

\begin{table}[!htp]
\caption{PBE0 spin constant matrix for Ce (in eV). Identical results were obtained for for Ce and CeO$_2$ parameterizations since the STO basis set remained unchanged for these calculations. }
\label{tab:ce_spin_constants_pbe0}
\begin{tabular}{c|cccc}
\hline
 & 4f & 5d & 6s & 6p \\
\hline
4f & -0.01885 & -0.00281 & -0.00082 & -0.00281 \\
5d & -0.00281 & -0.01000 & -0.00604 & -0.01540 \\
6s & -0.00082 & -0.00604 & -0.01249 & -0.00604 \\
6p & -0.00281 & -0.01540 & -0.00604 & -0.00625 \\
\hline
\end{tabular}
\end{table}

\begin{table}[!tbh]
\captionsetup{labelfont={color=black}}
\caption{\red{Decoding mask used in CeO$_2$ DSKO optimizations. All band positions are shown relative to the uppermost valence band, which is defined as `0'.}}
\label{tab:decoding mask}

{\color{black}
\begin{tabular}{cc}
\hline
Special k-point & Bands  \\
\hline
\hline
$\Gamma$ & $\left[-3, -2, -1, 0, 1, 4\right]$ \\
$X$ & $\left[-3, -2, -1, 0, 1\right]$ \\
$W$ & $\left[-3, -2, -1, 0, 1\right]$ \\
$K$ & $\left[-3, -2, -1, 0, 1\right]$ \\
$L$ & $\left[-3, -2, -1, 0, 1\right]$ \\
$U$ & $\left[-3, -2, -1, 0, 1\right]$ \\
\hline
\end{tabular}
}
\end{table}

\begin{table}[!htp]
\captionsetup{labelfont={color=black}}
\caption{\red{Confinement parameters and Hubbard $U$ values for Ce and O used in the Ce--O parametrization.}}
\label{tab:ce_o_params}

{\color{black}
\begin{tabular}{cccccc}
\hline
Element & Shell & $r_0$ & $a$ & $W$ & Hubbard $U$ (eV) \\
\hline
\hline
Ce & 4f & 4.0056 & 30.3612 & 5.2896 & 0.1645 \\
Ce & 5d & 9.5332 & 35.5609 & 3.9580 & 0.1645 \\
Ce & 6s & 7.6182 & 41.2194 & 3.5980 & 0.1645 \\
Ce & 6p & 7.4353 & 35.2704 & 2.3549 & 0.1645 \\
\hline
Ce & Density & 17.1335 & 43.4927 & 47.9009 & --- \\
\hline
O & 2s & 5.1296 & 25.5178 & 3.5042 & 0.5168 \\
O & 2p & 6.5569 & 49.4811 & 4.4037 & 0.5168 \\
\hline
O & Density & 16.0555 & 82.3198 & 20.9091 & --- \\
\hline
\end{tabular}
}
\end{table}

\begin{table}[!htp]
\captionsetup{labelfont={color=black}}
\caption{\red{PBE0 spin constant matrix for O used in the Ce--O parametrization (in eV).}}
\label{tab:o_spin_constants_pbe0}

{\color{black}
\begin{tabular}{c|cc}
\hline
 & 2s & 2p \\
\hline
2s & -0.06540 & -0.04334 \\
2p & -0.04334 & -0.03796 \\
\hline
\end{tabular}
}
\end{table}

\newpage
\section{\red{CeO$_2$ band structure orbital projections}}

\begin{figure}[!htp]
\centering
\includegraphics[scale=0.20]{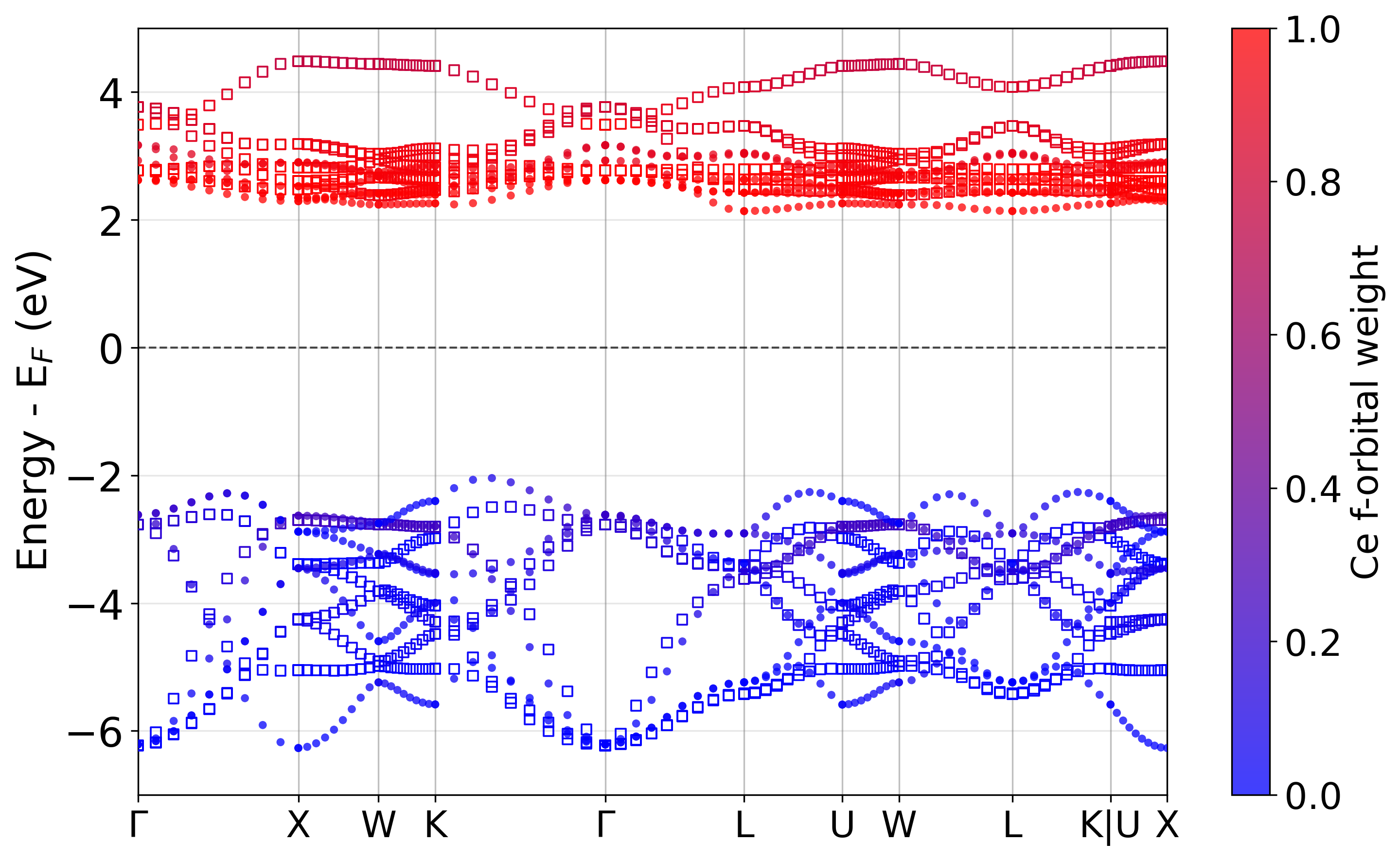}
\includegraphics[scale=0.20]{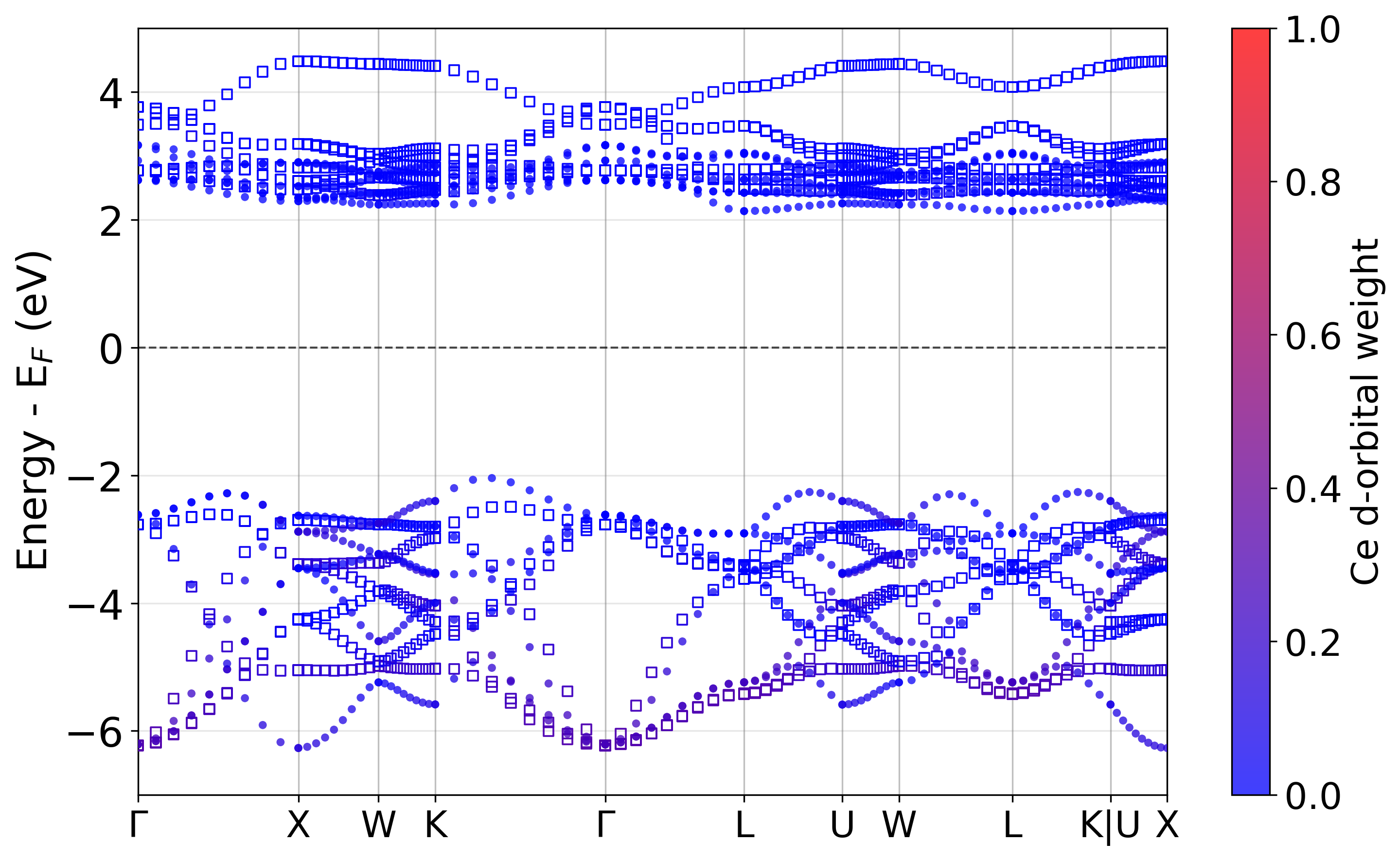}
\includegraphics[scale=0.20]{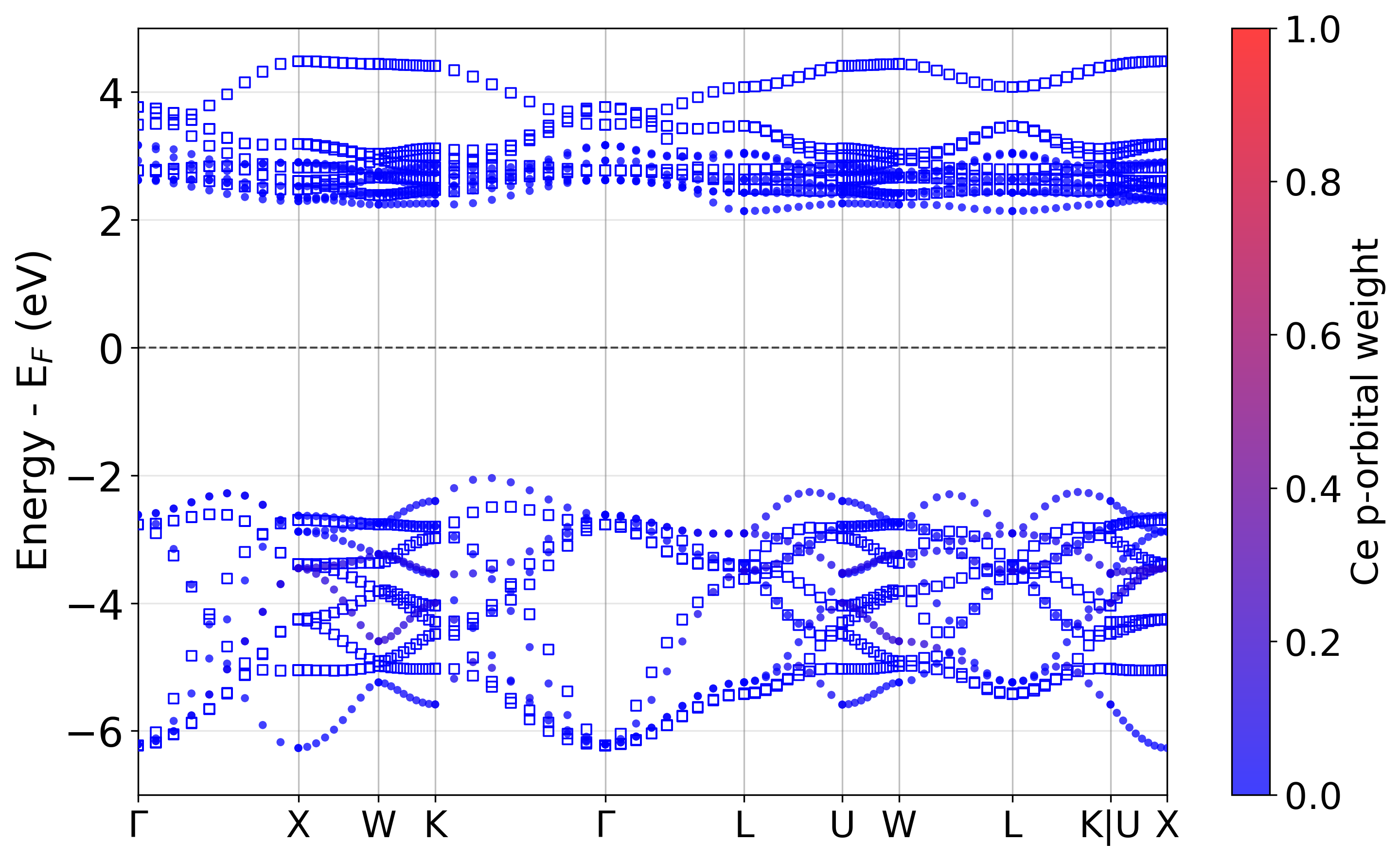}
\includegraphics[scale=0.20]{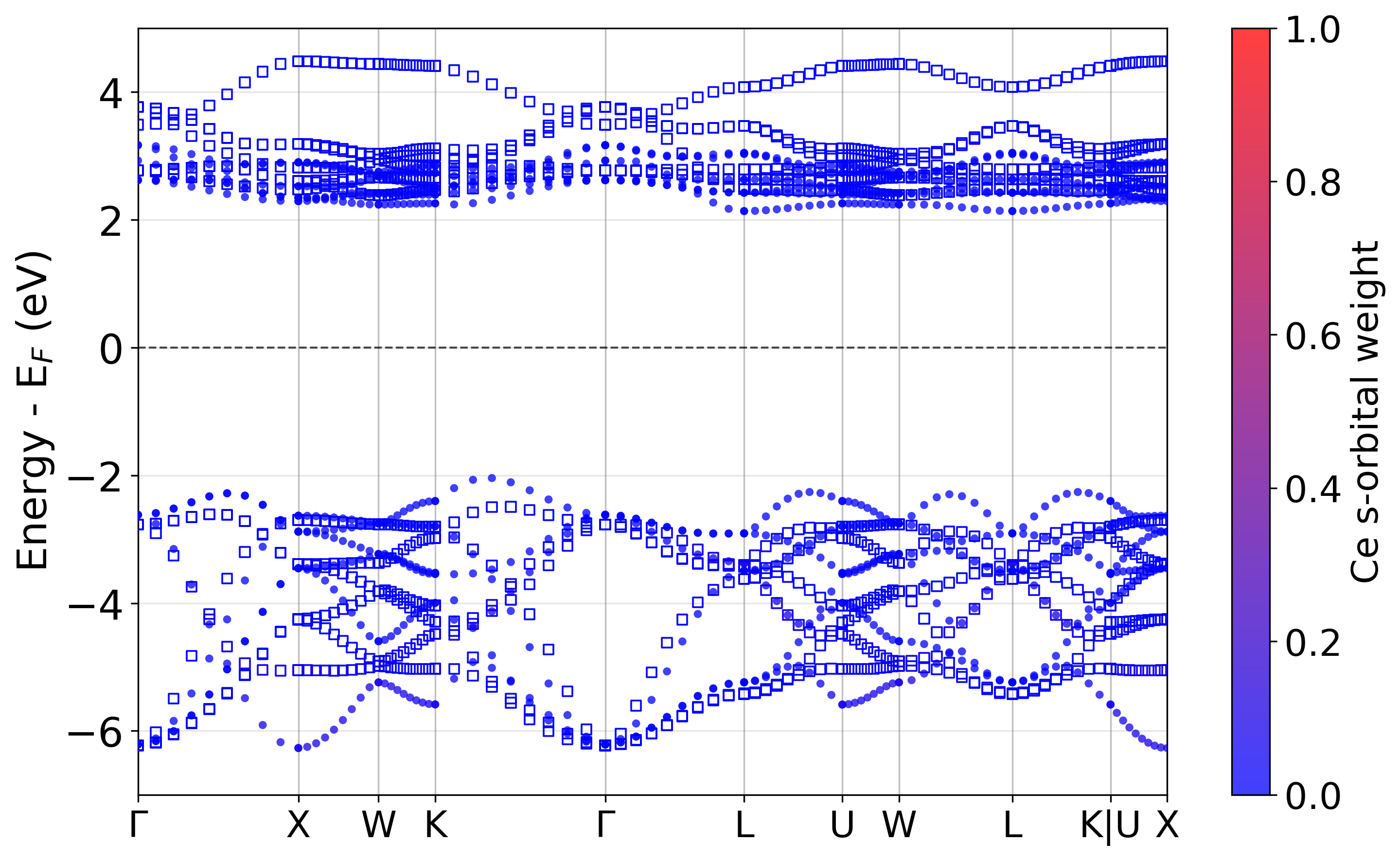}
\\
\includegraphics[scale=0.20]{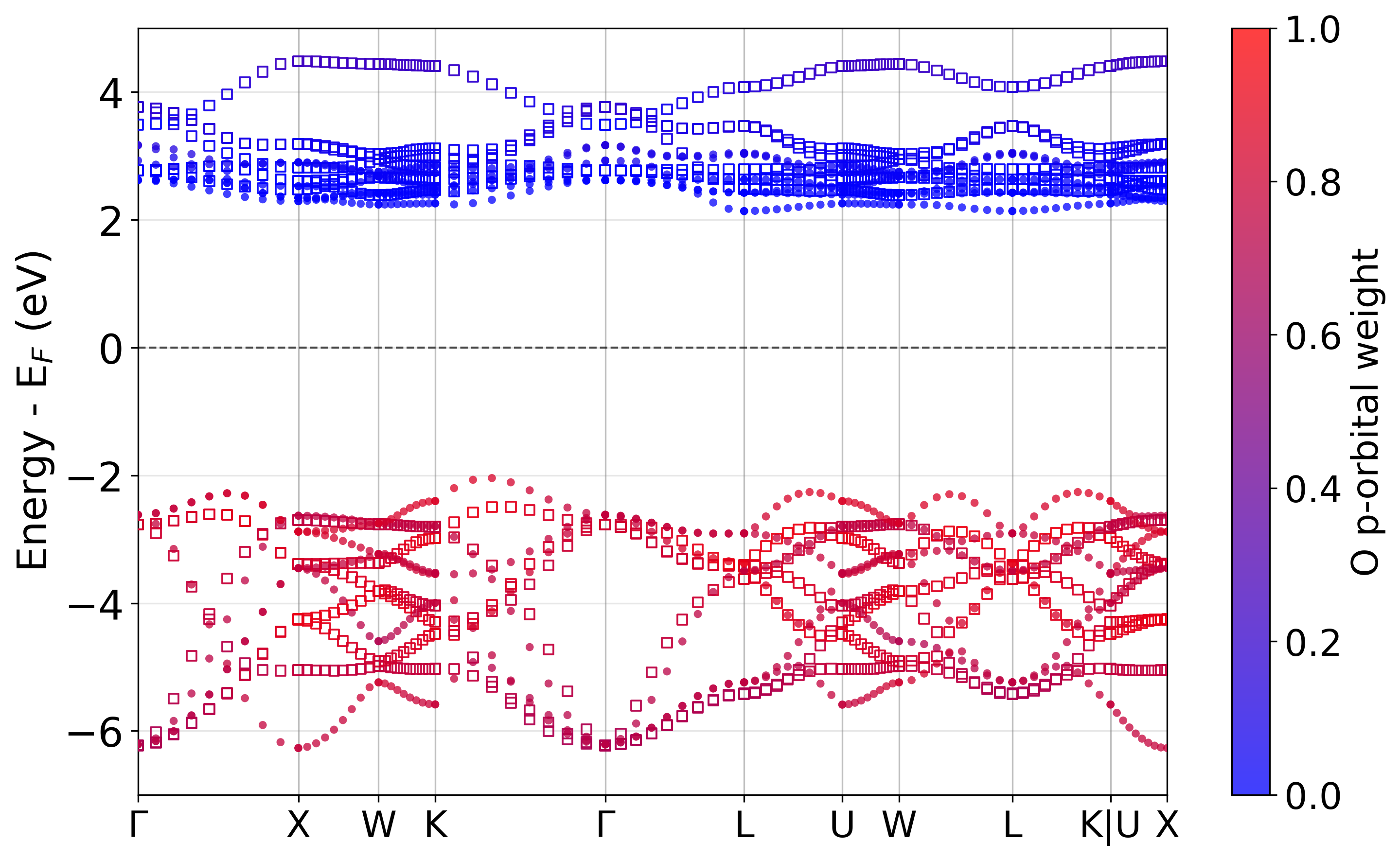}
\includegraphics[scale=0.20]{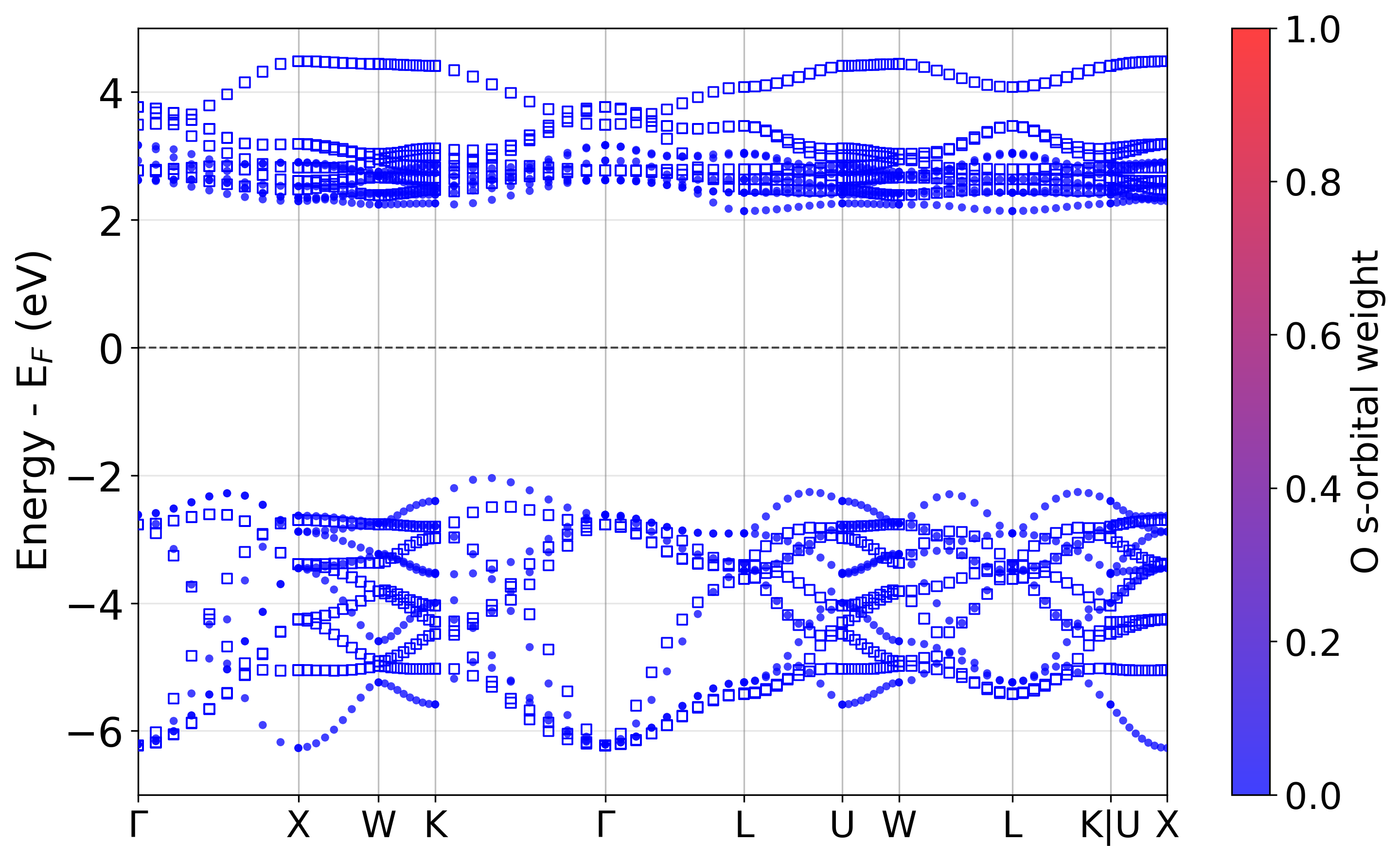}
\captionsetup{labelfont={color=black}} \caption{\label{fig:ceo2_pbe0} \red{Atom resolved band structure projections for CeO$_2$ from our DFTB-PBE0 model. Closed circles correspond to results from DFT and open squares to DFTB$+$. Red indicates increased orbital projection values.} }
\end{figure}

\newpage
\bibliography{library}